\documentclass[useAMS,usenatbib]{mn2e}
\usepackage[fleqn]{amsmath}
\usepackage{graphicx}
\usepackage{threeparttable}
\usepackage{booktabs}
\title[Locating Bound Structure]{Locating Bound Structure in an Accelerating Universe}
\author[D. W. Pearson and D. J. Batuski]{David W. Pearson$^{1}$\thanks{E-mail: david.pearson@umit.maine.edu} and David J. Batuski$^{1}$ \\
$^{1}$ Department of Physics and Astronomy, University of Maine, 120 Bennett Hall, Orono, ME 04469, USA}
\begin{document}

\date{Accepted 2013 August 23. Received 2013 August 22; in original form 2013 April 5}

\pagerange{\pageref{firstpage}--\pageref{lastpage}} \pubyear{2013}

\maketitle

\label{firstpage}

\begin{abstract}
Given the overwhelming evidence that the universe is currently undergoing an accelerated expansion, the question of what are the largest gravitationally bound structures remains. A couple of groups, \cite{Busha03} (B03) and \cite{Dunner06} (D06), have attempted to analytically define these limits, arriving at substantially different estimates due to differences in their assumptions about the velocities at the present epoch. In an effort to locate the largest bound structures in the universe, we selected the Aquarius (ASC), Microscopium (MSC), Corona Borealis (CBSC), and Shapley (SSC) superclusters for study, due to their high number density of rich Abell clusters. Simple $N$-body simulations, which assumed negligible intercluster mass, were used to assess the likelihood of these structures being gravitationally bound, and the predictions of the  models of B03 and D06 were compared with those results. We find that ASC, and MSC contain pairs of clusters which are gravitationally bound, A2541/A2546 and A3695/A3696 respectively, with no other structures having a significant chance of being bound. For SSC, we find a group of five clusters, A3554, A3556, A3558, A3560, and A3562 that are bound, with an additional pair, A1736/A3559, having a slight chance of being bound. We find that CBSC has no extended bound structure, contrary to the findings of \cite{Small98}, who claim that the entire supercluster is bound. In regards to the analytical models, we find that B03 will identify structure that is definitely bound, but tends to underestimate the true extent of the structure, while D06 will identify all structure that is bound while overestimating its extent. Combined, the two models can provide lower and upper limits to the extent of bound structures so long as there are no other significant structures nearby or no significant dark matter exterior to the clusters.
\end{abstract}


\section{Introduction}
Observations of type Ia supernovae \citep{Perlmutter99,Amanullah10} and the cosmic microwave background \citep{Spergel03,Ade13} have shown that the universe is currently in a phase of accelerating expansion. The simplest model consistent with the data is the $\Lambda$CDM model, consisting primarily of a cosmological constant dark energy component and cold dark matter. In such a universe, once the accelerated expansion stage has begun, large scale structure formation will be strongly suppressed, leaving only existing bound structures to collapse. The question of the extent of those structures is difficult to answer. Two groups, \cite{Busha03} (B03) and \cite{Dunner06} (D06), have attempted to define the limits of bound structures using the spherical collapse model, but arrive at fairly different answers. Upon comparing with the results of some $N$-body simulations, neither group claims their model is completely consistent with the simulation results. B03 attribute this to their assumption that particles will currently be streaming away from structures with an unperturbed Hubble flow velocity, which is probably not true for gravitationally bound structures. D06, on the other hand, claims the failings of their model are due to ignoring tangential motions of particles and external attractors.

\cite{Dunner07} go further, by extrapolating the model of D06 to define redshift-space limits of bound structure. They develop three methods for fitting velocity envelopes to redshift data to identify bound structure. The errors in bound mass from their methods are $\sim 30$--$40$\% mainly attributed to the morphology and substructure present in the object being studied.

In order to locate the largest gravitationally bound structures in the universe, we selected four superclusters for study due to their high number densities of rich Abell clusters. The Aquarius (ASC), Microscopium (MSC), Corona Borealis (CBSC), and Shapley (SSC) superclusters have number densities that are 40 to over 100 times the average number density of rich Abell clusters, the highest known in the local universe (\mbox{$z \la 0.2$}). This gives these superclusters the best chance of having extended bound structures.

This paper is structured as follows. Section 2 discusses the physical properties of the superclusters, including the masses of the individual clusters. Section 3 details the $N$-body simulations that were performed to determine the likelihood that the superclusters are gravitationally bound. Section 4 presents the results of the simulations. Section 5 offers some discussion on the implications of those results.
\section{The Supercluster Candidates}
\subsection{Previous Work}
Other groups have studied these superclusters in varying levels of detail. SSC ($z \sim 0.05$) has been studied by numerous authors in a variety of different wavelengths, since it is the densest concentration of galaxies and galaxy clusters in the local universe. \cite{Hanami99} examined the x-ray properties of the members of the core region made up of A3556, A3558, A3562, SC 1329-313, and SC 1327-312. They find that SC 1329-313 is likely an object currently undergoing a merger. They explain A3556 as a post-merger object where the intracluster gas was heated and expanded adiabatically for $\sim 10^{9} \, \mathrm{yr}$, giving it the low X-ray luminosity they observed. They conclude that the other clusters have likely formed as the result of more recent mergers. \cite{Reisenegger00} applied the spherical collapse model to the core region to define velocity caustics which separate the collapsing structure from the foreground and background galaxies. From this they determine that the inner 8 $h^{-1} \, \mathrm{Mpc}$ ($H_{0} = 100 \, h \, \mathrm{km} \, \mathrm{s}^{-1} \, \mathrm{Mpc}^{-1}$) core is currently collapsing, that the current turn around radius is $\sim 14 \, h^{-1} \, \mathrm{Mpc}$, and that the bound region is $\sim 20 \, h^{-1} \, \mathrm{Mpc}$. \cite{Proust06} performed a large survey of the region and find evidence for the intercluster galaxies contributing twice as much mass as the cluster galaxies, allowing SSC to have a non-negligible effect on the Local Group's peculiar motion. This would also seem to indicate a great possibility of extending the size of the bound region of this supercluster.

CBSC ($z \sim 0.07$) has been studied by fewer authors than SSC. \cite{Postman88} obtained mass estimates for A2061, A2065, A2067, A2079, A2089 and A2092. They combine these with an estimate of the intercluster mass extrapolated from five fields using an assumed mass to light ratio to arrive at a supercluster mass of $\sim 8.2 \times 10^{15} \, \mathrm{M}_{\odot}$. Due to an uncertainty of about a factor of two, this mass would be consistent with the binding mass ($M_{\mathrm{binding}} = M_{\mathrm{virial}}/2 = 1.3 \times 10^{16} \mathrm{M}_{\odot}$). Using the data from the Norris Survey, \cite{Small98} applied the Virial theorem on the supercluster scale to estimate the mass of CBSC. They arrived at an estimate of $\sim 4\times 10^{16} \, h^{-1} \, \mathrm{M}_{\odot}$, which would of course be enough mass to bind CBSC. \cite{Kopylova98} applied the Kormendy relation \citep{Kormendy77} to the supercluster to get redshift independent distance estimates for the clusters. Using the distance estimates and spectroscopic redshifts they concluded that the supercluster displays signs of being in a state of collapse. More recently, Batiste \& Batuski (2013, in preparation) have found similar results using the fundamental plane \citep{Djorgovski87}.

ASC ($z \sim 0.11$) has only been studied by a couple of groups. \cite{Batuski99} first noted a significant concentration of six Abell clusters in the Aquarius region. They were unable to come to firm conclusions about the dynamics given that they had no secondary distance indicators and, for some clusters, they only had a few redshifts. \cite{Caretta02,Caretta04} performed a more detailed study of the region, obtaining more redshift information, and were able to get better estimates of redshifts and velocity dispersions of the clusters. They estimate the mass of the supercluster by summing the estimated masses of 14 clusters spread over a large filamentary volume, arriving at $\sim 2 \times 10^{16} \, h^{-1} \, \mathrm{M}_{\odot}$. Neither group made any firm claims about the densest part of this structure being gravitationally bound.

\cite{Batuski99} also noted that MSC ($z \sim 0.09$) could potentially be in this category of high density regions, adding a seventh cluster to the six found by \cite{Zucca93}. Using redshifts from \cite{Katgert96}, \cite{Batuski99} noted the potential for an overdensity of about 130 times the average number density of Abell clusters. No researchers have published any attempts to explore the dynamics of this supercluster, leaving the extent of potentially bound structure unknown.
\subsection{Supercluster Properties}
\begin{table*}
\caption{Cluster Properties}
\label{ClusterProp}
\centering
\begin{threeparttable}
\begin{tabular}{cccccccc}
\hline
\hline
\rule{0pt}{12pt}
Supercluster & Cluster & RA\tnote{a} & Dec\tnote{a} & $N$ & $M$ & $z$ & References\tnote{b} \\
${}$ & ${}$ & (J2000) & (J2000) & ${}$ & ($10^{15} \, h^{-1} \, \mathrm{M}_{\odot}$) & ${}$ & ${}$ \\[1ex]
\hline
\rule{0pt}{12pt}
ASC & A2541 & $23^{\mathrm{h}}10^{\mathrm{m}}04^{\mathrm{s}}$ & $-22^{\circ}57'43''$ & 15 & $0.591\pm 0.084$ & 0.1135 & 1,4,9 \\
${}$ & A2546 & $23^{\mathrm{h}}10^{\mathrm{m}}46^{\mathrm{s}}$ & $-22^{\circ}39'42''$ & 21 & $0.422\pm 0.055$ & 0.1131 & 1,4,9 \\
${}$ & A2548 & $23^{\mathrm{h}}11^{\mathrm{m}}21^{\mathrm{s}}$ & $-20^{\circ}25'41''$ & 6 & $0.005\pm 0.007$\tnote{c} & 0.1104 & 1,4,9 \\
${}$ & A2554 & $23^{\mathrm{h}}12^{\mathrm{m}}15^{\mathrm{s}}$ & $-21^{\circ}33'55''$ & 33 & $0.660\pm 0.074$ & 0.1109 & 1,3,4,9 \\
${}$ & A2555 & $23^{\mathrm{h}}12^{\mathrm{m}}45^{\mathrm{s}}$ & $-22^{\circ}12'40''$ & 9 & $0.043\pm 0.010$ & 0.1109 & 1,4,9 \\
${}$ & A3985 & $23^{\mathrm{h}}15^{\mathrm{m}}57^{\mathrm{s}}$ & $-23^{\circ}19'37''$ & 8 & $0.173\pm 0.070$ & 0.1132 & 1,5,9 \\ [1ex]
\hline
\rule{0pt}{12pt}
MSC & A3677 & $20^{\mathrm{h}}26^{\mathrm{m}}21^{\mathrm{s}}$ & $-33^{\circ}21'06''$ & 7 & $0.921\pm 0.189$ & 0.091 & 4,6,8 \\
${}$ & A3693 & $20^{\mathrm{h}}34^{\mathrm{m}}22^{\mathrm{s}}$ & $-34^{\circ}29'40''$ & 11 & $0.871\pm 0.237$ & 0.091 & 6,9,10 \\
${}$ & A3695 & $20^{\mathrm{h}}34^{\mathrm{m}}48^{\mathrm{s}}$ & $-35^{\circ}49'39''$ & 48 & $0.652\pm 0.241$ & 0.089 & 4,6,8,9 \\
${}$ & A3696 & $20^{\mathrm{h}}35^{\mathrm{m}}10^{\mathrm{s}}$ & $-34^{\circ}54'36''$ & 12 & $0.023\pm 0.014$ & 0.088 & 6,9 \\
${}$ & A3705 & $20^{\mathrm{h}}41^{\mathrm{m}}42^{\mathrm{s}}$ & $-35^{\circ}14'00''$ & 30 & $0.142\pm 0.113$ & 0.090 & 3,4,6,9 \\ [1ex]
\hline
\rule{0pt}{12pt}
CBSC & A2061 & $15^{\mathrm{h}}21^{\mathrm{m}}21^{\mathrm{s}}$ & $30^{\circ}40'15''$ & 105 & $1.708\pm 0.159$ & 0.0772 & 2,7 \\
${}$ & A2065 & $15^{\mathrm{h}}22^{\mathrm{m}}43^{\mathrm{s}}$ & $27^{\circ}43'21''$ & 27 & $0.951\pm 0.182$ & 0.0714 & 2,7 \\
${}$ & A2067 & $15^{\mathrm{h}}23^{\mathrm{m}}14^{\mathrm{s}}$ & $30^{\circ}54'24''$ & 50 & $1.184\pm 0.133$ & 0.0736 & 2,7 \\
${}$ & A2089 & $15^{\mathrm{h}}32^{\mathrm{m}}45^{\mathrm{s}}$ & $28^{\circ}03'47''$ & 23 & $0.361\pm 0.062$ & 0.0720 & 2,7 \\ [1ex]
\hline
\rule{0pt}{12pt}
SSC & A1736 & $13^{\mathrm{h}}26^{\mathrm{m}}52^{\mathrm{s}}$ & $-27^{\circ}06'33''$ & 87 & $0.696\pm 0.067$ & 0.0446 & 3,4,7,8,9,11,12,13,14 \\
${}$ & A3528 & $12^{\mathrm{h}}54^{\mathrm{m}}18^{\mathrm{s}}$ & $-29^{\circ}01'16''$ & 43 & $0.309\pm 0.037$ & 0.0516 & 3,4,7,8,9,11,12 \\
${}$ & A3530 & $12^{\mathrm{h}}55^{\mathrm{m}}36^{\mathrm{s}}$ & $-30^{\circ}21'12''$ & 77 & $0.692\pm 0.033$ & 0.0525 & 3,4,7,8,9,11,12 \\
${}$ & A3532 & $12^{\mathrm{h}}57^{\mathrm{m}}19^{\mathrm{s}}$ & $-30^{\circ}22'13''$ & 82 & $0.558\pm 0.042$ & 0.0542 & 3,4,7,8,9,11,12,13 \\
${}$ & A3542 & $13^{\mathrm{h}}08^{\mathrm{m}}41^{\mathrm{s}}$ & $-34^{\circ}34'00''$ & 10 & $0.086\pm 0.014$ & 0.0513 & 3,4,7,8,9,11,12 \\
${}$ & A3553 & $13^{\mathrm{h}}19^{\mathrm{m}}14^{\mathrm{s}}$ & $-37^{\circ}10'48''$ & 26 & $0.148\pm 0.029$ & 0.0475 & 3,4,7,8,9,11,12,13 \\
${}$ & A3554 & $13^{\mathrm{h}}19^{\mathrm{m}}30^{\mathrm{s}}$ & $-33^{\circ}28'42''$ & 32 & $0.260\pm 0.025$ & 0.0458 & 3,4,7,8,9,11,12,13 \\
${}$ & A3556 & $13^{\mathrm{h}}24^{\mathrm{m}}06^{\mathrm{s}}$ & $-31^{\circ}39'38''$ & 58 & $0.520\pm 0.078$ & 0.0467 & 3,4,7,8,9,11,12 \\
${}$ & A3558 & $13^{\mathrm{h}}27^{\mathrm{m}}55^{\mathrm{s}}$ & $-31^{\circ}29'32''$ & 205 & $0.931\pm 0.050$ & 0.0468 & 3,4,7,8,9,11,12,13 \\
${}$ & A3559 & $13^{\mathrm{h}}29^{\mathrm{m}}54^{\mathrm{s}}$ & $-29^{\circ}31'29''$ & 41 & $0.647\pm 0.053$ & 0.0449 & 3,4,7,8,9,11,12 \\
${}$ & A3560 & $13^{\mathrm{h}}32^{\mathrm{m}}24^{\mathrm{s}}$ & $-33^{\circ}05'24''$ & 50 & $0.804\pm 0.090$ & 0.0477 & 3,4,7,8,9,11,12,13 \\
${}$ & A3562 & $13^{\mathrm{h}}33^{\mathrm{m}}32^{\mathrm{s}}$ & $-31^{\circ}40'23''$ & 76 & $1.102\pm 0.102$ & 0.0478 & 3,4,7,8,9,11,12,13 \\
${}$ & A3564 & $13^{\mathrm{h}}34^{\mathrm{m}}22^{\mathrm{s}}$ & $-35^{\circ}13'24''$ & 22 & $0.315\pm 0.024$ & 0.0493 & 4,7,9 \\
${}$ & A3566 & $13^{\mathrm{h}}38^{\mathrm{m}}59^{\mathrm{s}}$ & $-35^{\circ}33'12''$ & 42 & $0.289\pm 0.070$ & 0.0498 & 3,4,7,8,9,11,12 \\
${}$ & A3577 & $13^{\mathrm{h}}54^{\mathrm{m}}20^{\mathrm{s}}$ & $-27^{\circ}50'42''$ & 32 & $0.274\pm 0.021$ & 0.0482 & 3,5,7,8,9,10,11,12,13 \\ [1ex]
\hline
\end{tabular}
\begin{tablenotes}
\item[a] Right ascension and declination from \cite{ACO}
\item[b] \emph{References:} (1) \cite{Caretta04}, (2) \cite{Small97}, (3) \cite{Fairall91}, (4) \cite{CfA}, (5) \cite{2dF}, (6) \cite{Katgert96}, (7) \cite{Struble99}, (8) \cite{Wang09}, (9) \cite{6dF}, (10) \cite{Guzzo09}, (11) \cite{Lavaux11}, (12) \cite{Huchra12}, (13) \cite{Dressler91}, (14) \cite{DaCosta98}
\item[c] Given the small number of redshifts for this cluster, it is likely to contribute more mass than what is quoted here.
\end{tablenotes}
\end{threeparttable}
\end{table*}
We began by identifying the highest peaks in the number density of rich Abell clusters in order to identify candidates for gravitationally bound superclusters. To do this we found the apparent three dimensional positions using redshift as a distance indicator \citep{Sandage75}
\begin{equation}
\label{distance}
D = \dfrac{cz}{H_{0}} \dfrac{(1+z/2)}{1+z}.
\end{equation}
We then fit spheres just large enough to encompass the most distant cluster to estimate the number density.  These densities were compared to the average Abell, Corwin, Olowin (ACO) cluster density of $8 \times 10^{-6} \, h^{3} \, \mathrm{Mpc}^{-3}$ \citep{Miller99}. Different combinations of the clusters associated with the various superclusters were tested until we found a sufficiently dense peak. ASC, MSC, CBSC, and SSC were selected as candidates for gravitationally bound superclusters due to having number densities of about 130, 110, 70, and 40 times the average number density respectively. 

It is worth noting that SSC has a core of five clusters which has an overdensity of about 690 times the average. The smaller number density of 40 times the average includes a total of 15 clusters contained within a sphere of roughly $22 \, h^{-1} \, \mathrm{Mpc}$, approximately the size of the bound region claimed by \cite{Reisenegger00}. If A3558 does represent the location of the core, then 12 clusters would be contained in a sphere of roughly $18 \, h^{-1} \, \mathrm{Mpc}$ (with the other 3 clusters being more than $20 \, h^{-1} \, \mathrm{Mpc}$ away from A3558) yielding an overdensity of about 60 times the average.

The masses of the individual clusters were needed in order to perform the simulations. These were determined by searching the National Virtual Observatory (NVO) archive for all available redshifts within a projected $1.5 \, h^{-1} \, \mathrm{Mpc}$ radius of the cluster centre, and combining them into a single catalogue. We then removed duplicate entries, and foreground and background galaxies as determined through a histogram analysis. Finally, repeated measurements of the same galaxy were averaged together. Once a final list of galaxies was obtained, we estimated the mass using the methods of \cite{Carlberg96}. First we found the dispersion,
\begin{equation}
\label{dispersion}
\sigma^{2} = \left(\sum_{i} w_{i}\right)^{-1} \sum_{i} w_{i} \left(\Delta v_{i}\right)^{2},
\end{equation}
where $\Delta v_{i}$ is the difference of the $i^{\mathrm{th}}$ galaxies redshift with the average cluster redshift, and $w_{i}$ is the weight. Next the projected mean harmonic separation was calculated,
\begin{equation}
\label{harmonic}
R_{H}^{-1} = \left(\sum_{i} w_{i}\right)^{-2} \sum_{i<j} \dfrac{w_{i}w_{j}}{|r_{i}-r_{j}|},
\end{equation}
where $r_{i}$ is the position on the sky. The virial mass is then
\begin{equation}
\label{mvirial}
M_{v} = \dfrac{3\pi \sigma^{2} R_{H}}{2G}.
\end{equation}

Error estimates in the calculated masses for ASC, MSC and CBSC come from jackknife resampling \citep{Beers90} of the virial mass. When this procedure was applied to SSC, the error estimates were relatively small, likely due to the more extensive and uniform sampling of the clusters on the sky. For this reason the jackknife was applied to the dispersion and harmonic radius instead, and those errors were then added in quadrature to give the errors for SSC. Since we use these errors to test the robustness of our results (see section 3.1), having larger error estimates, similar in magnitude to those for the other superclusters, was desirable. Table \ref{ClusterProp} lists the cluster properties for all clusters used in the various simulations. Column 1 identifies the supercluster to which the clusters belong. Column 2 gives the Abell cluster number. Columns 3 and 4 give the right ascension and declination in J2000 coordinates respectively. Column 5 gives the number of galaxies used in obtaining the mass estimates. Column 6 gives the mass estimate found from equations \eqref{dispersion}, \eqref{harmonic}, and \eqref{mvirial} along with the 68 per cent confidence limits found from the statistical jackknife. Column 7 lists the cluster redshifts. Lastly, column 8 lists the references used for redshifts and positions of each cluster.
\section{N-Body Simulations}
In order to assess the likelihood of the superclusters being gravitationally bound, we ran a series of $N$-body simulations. Since the internal structure of the clusters was not of interest, we developed our own simulation software which modelled them as point particles. This greatly decreased the computation time per simulation allowing us to explore more realisations of the initial conditions. This was also the reason for developing new simulation software instead of employing an available system like GADGET2 \citep{Springel05}. The lack of complexity needed, and small number of particles would make using a system like GADGET2 more computationally expensive, without gaining useful accuracy. The cosmological parameters used were $\Omega_{m,0} = 0.26$, $\Omega_{\Lambda,0} = 0.74$, $h = 0.74$, and $H_{0} = 74 \, \mathrm{km} \, \mathrm{s}^{-1} \, \mathrm{Mpc}^{-1}$. 

For the simulations, we only included the calculated masses of the clusters themselves. The primary reasons for this were lack of conclusive evidence of a significant mass component outside of the clusters, and lack of a reliable way of estimating how much mass is outside of the clusters. \cite{Proust06} find that in SSC the intercluster galaxies may contribute about twice as much mass to the supercluster as the cluster galaxies. However, SSC represents an extreme overdensity, making it difficult to justify making that assumption for all of the superclusters in our study. Considering that intercluster galaxies are likely going to be more homogeneously distributed over the supercluster volume, their effect should be a fairly uniform lowering of the gravitational potential while the rich clusters represent deep potential wells. Given this, we assumed that the intercluster galaxies would mostly act as tracers of the dynamics of the supercluster, which should be dominated by the interactions of the clusters.

As for dark matter exterior to the clusters, there is only some contradictory evidence for filaments between pairs of clusters at present.  \cite{Gray02} found evidence for a dark matter filament between A901 and A902 from a ground based weak lensing survey. \cite{Heymans08} did a follow up study of the pair with the Hubble space telescope (HST) STAGES survey data. They do not recover the signal of this filament and attribute the original detection to residual point spread function systematics and the type of mass reconstruction used by \cite{Gray02}. \cite{Dietrich12} find evidence of a dark matter filament between A222 and A223 from a weak lensing study which seems to be backed by a coincidence of a galaxy overdensity and diffuse, soft-X-ray emission. Given this lack of conclusive evidence on filaments of dark matter between clusters, we chose not to include those effects in our initial simulation runs. Future work will be aimed at exploring the effects of intercluster matter in detail (Pearson \& Batuski, in preparation).
\subsection{Initial Conditions}
\label{ICs}
The initial conditions for the simulations were based on the available observational data (positions on the sky and redshifts). From these data, three dimensional positions were calculated using the redshift as a distance indicator via equation \eqref{distance}. Initial velocities were assigned as the combination of two components, a Hubble flow component and a randomly oriented circular orbit velocity around the combined centre of mass of the other clusters in the supercluster.

These initial conditions were then varied so that the distributions of the varied quantities would be Gaussian. This was achieved by creating a Gaussian distribution based on the input value as the mean, with a $\sigma$ value corresponding to the error estimates in the quantity. A value was then selected randomly from that distribution. To vary the positions, the redshifts of the clusters were fed into the algorithm described above, varying the line-of-sight distance from Earth. The velocities were varied by feeding each Cartesian component into the algorithm. In this manner 250 sets of initial conditions per supercluster were generated.

To gain a sense of how the uncertainty in mass may effect our results, we performed Gaussian random mass realisations. Using the uncertainties listed in Table \ref{ClusterProp} as one standard deviation, a Gaussian distribution of each clusters mass was generated. A single mass realisation then consists of randomly drawing a mass for each cluster in the supercluster from these distributions. For each mass realisation, simulations were run based on all 250 realisations of the initial conditions. In order to keep the total number of simulations to a reasonable quantity, the number of mass realisations was capped at 20. For each supercluster we then ran a total of 5,250 simulations (250 using the mass central values and 250 per mass realisations).

\subsection{Physics}
The simulations included the effects of gravity and cosmological constant dark energy. Gravity was modelled using a spline softening technique \citep{Hernquist89} where the acceleration of the $i^{\mathrm{th}}$ particle is given as
\begin{equation}
\label{agrav}
\bmath{a}_{g,i} = \sum_{j \neq i}^{N} Gm_{j}g_{ij}\bmath{r}_{ij}.
\end{equation}
Here, $G$ is Newton's gravitational constant, $m_{j}$ is the mass of the $j^{\mathrm{th}}$ particle, $\mathbf{r}_{ij}$ is the separation of the $i^{\mathrm{th}}$ and $j^{\mathrm{th}}$ particles, and $g_{ij}$ is given by 
\begin{equation}
\label{g}
g(r) = \left\{\begin{array}{lr}
\frac{1}{\varepsilon^{3}}\left[\frac{4}{3}-\frac{6u^{2}}{5}+\frac{u^{3}}{2}\right] & 0 \leq u\leq 1 \\
\frac{1}{r^{3}}\left[-\frac{1}{15}+\frac{8u^{3}}{3}-3u^{4}+\frac{6u^{5}}{5}-\frac{u^{6}}{6}\right] & 1 \leq u \leq 2 \\
\frac{1}{r^{3}} & u \geq 2
\end{array}\right. ,
\end{equation}
where $\varepsilon$ is the softening length ($1.5 \, \mathrm{Mpc}$), and $u = r/\varepsilon$. In this way, gravity behaves as completely Newtonian for any distance beyond two softening lengths, and then smoothly turns over and goes to zero as two particles approach the same location.

The effects of the cosmological constant dark energy were, in essence, modelled as a gravitational force as well. \cite{Baryshev01} give the gravitating mass of any component of the universe as
\begin{equation}
\label{gravmass}
M_{Q} = \dfrac{4\pi}{3} \left(1+3w\right) \rho_{Q} r^{3}.
\end{equation}
where $w$ is the equation of state parameter, $\rho_{Q}$ is the matter density of the component of interest, which is equal to its energy density, $\epsilon_{Q}$, divided by the speed of light squared, and $r$ is the distance from the origin. Given that the energy density of cosmological constant dark energy is \citep{Ryden03}
\begin{equation}
\label{elambda}
\epsilon_{\Lambda} \equiv \dfrac{c^2}{8 \pi G} \Lambda,
\end{equation}
and with $w = -1$, the gravitating mass can be found to be
\begin{equation}
\label{mlambda}
M_{\Lambda} = -\dfrac{\Lambda r^{3}}{3G}.
\end{equation}
Plugging this mass into the equation for gravitational acceleration, we arrive at the acceleration due to cosmological constant dark energy
\begin{equation}
\label{alambda}
\bmath{a}_{\Lambda,i} = \dfrac{1}{3} \Lambda \bmath{r}_{i}.
\end{equation}
This allows the effects of dark energy to be included by simply adding an additional acceleration directed outwards from the origin, which we take to be the centre of mass of the supercluster for convenience.

\begin{figure*}
\centering
\includegraphics{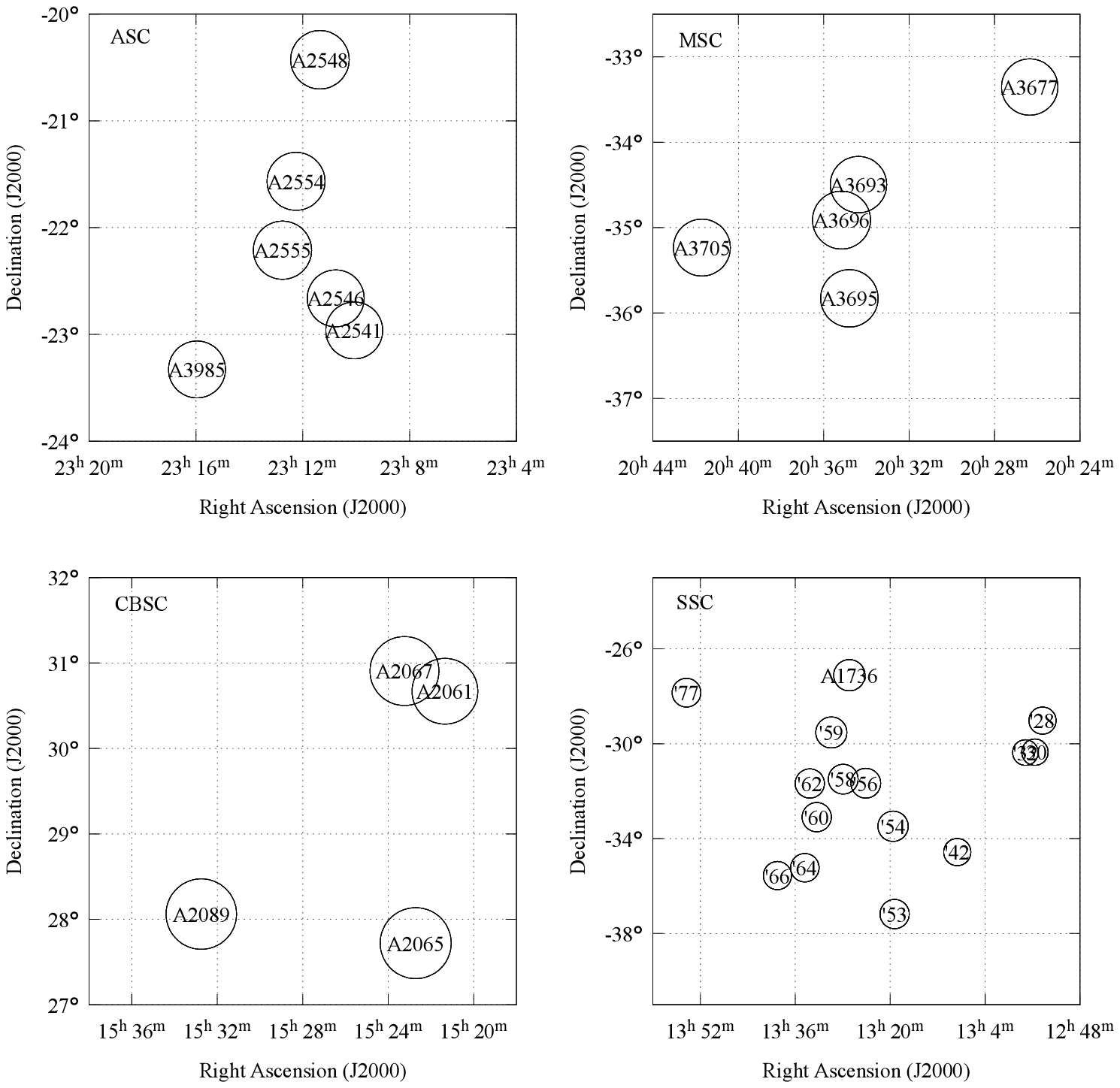}
\caption{Appearance of superclusters on the sky. The top left panel is ASC, which covers a $4^{\circ} \times 4^{\circ}$ region of the sky. The top right is MSC, which covers a $5^{\circ} \times 5^{\circ}$ region. The bottom left is CBSC, which covers a $5^{\circ} \times 5^{\circ}$ region. The bottom right is SSC, which covers a $18^{\circ} \times 18^{\circ}$ region. The circles represent a radius of $1.5 \, h^{-1} \, \mathrm{Mpc}$ around the centres of the clusters. For SSC all Abell numbers begin with A35- aside from A1736. On the right at about $-30^{\circ}$ declination, the two clusters that are obscuring each other are A3532 and A3530 from left to right.}
\label{SCSky}
\end{figure*}

\subsection{Integration}
In order to facilitate the integration of the simulation, a velocity Verlet algorithm was used. At each time step, first the particle positions are updated as
\begin{equation}
\label{position}
\bmath{r}_{i}(t+\delta t) = \bmath{r}_{i}(t) + \bmath{v}_{i}(t) \delta t + \dfrac{1}{2}\bmath{a}_{i}(t) \delta t^2.
\end{equation}
The velocities are at the same time partially updated as
\begin{equation}
\label{vel1}
\bmath{v}_{i}(t+\delta t/2) = \bmath{v}_{i}(t)+\dfrac{1}{2}\bmath{a}_{i}(t) \delta t.
\end{equation}
New accelerations are then calculated using the updated positions in equations \eqref{agrav}, \eqref{g} and \eqref{alambda}. Lastly, the velocity update is finished by the additional step of
\begin{equation}
\label{vel2}
\bmath{v}_{i}(t+\delta t) = \bmath{v}_{i}(t+\delta t/2)+\dfrac{1}{2}\bmath{a}_{i}(t+\delta t) \delta t.
\end{equation}

We make the distinction between two different time increments; the time step ($10^{14} \, \mathrm{s}$), and the calculation step, $\delta t$. After each time step the current positions, velocities and accelerations for the particles are output to files.  The calculation step is adaptive, based on the largest current acceleration 
\begin{equation}
\label{calcstep}
\delta t = 0.003e^{-60a_{\mathrm{max}}}.
\end{equation}
The coefficients were chosen so that energy and angular momentum were conserved over the course of the simulations. The code integrates by updating the positions, velocities, and accelerations over calculation steps, keeping track of the amount of time that has passed. If the next calculation step would take the simulation beyond a time step, it is reduced so that after that calculation step, one time step will be completed.

The simulations were first integrated backwards for 2763 time steps (8.76 Gyr) to test that the initial conditions were realistic, then forward for 11600 time steps (36.8 Gyr), or 8837 time steps (28 Gyr) from the present. When the forward integration phase reached the present time, the positions and velocities of the clusters were compared to the initial conditions to ensure that numerical error was negligible.

\subsection{Analysis}
\label{Analysis}
The output of the simulations was analysed to determine what portion of the superclusters were potentially gravitationally bound. In addition, the analytical models of B03 and D06 were tested to determine their predictive abilities. The analytical model of B03 predicts that anything within a radius $r_{0}$ of an object of mass $M_{\mathrm{obj}}$, will remain bound, with $r_{0}$ given by
\begin{equation}
\label{r_B03}
\dfrac{M_{\mathrm{obj}}}{10^{12} \, \mathrm{M}_{\odot}} \geq 3 \, h_{70}^{2} \left( \dfrac{r_{0}}{1 \, \mathrm{Mpc}}\right)^{3},
\end{equation}
where $h_{70} = H_{0}/70 \, \mathrm{km} \, \mathrm{s}^{-1} \, \mathrm{Mpc}^{-1}$. Similarly the prediction of D06 is
\begin{equation}
\label{r_D06}
\dfrac{M_{\mathrm{obj}}}{10^{12} \, \mathrm{M}_{\odot}} \geq 1.18 \, h_{70}^{2} \left( \dfrac{r_{0}}{1 \, \mathrm{Mpc}}\right)^{3}.
\end{equation}
Both of the equations above give estimates of the extent of bound structure in the present day universe. With these equations and the masses listed in Table \ref{ClusterProp}, each cluster's $r_{0}$ value was calculated. Based on the particular realization of the initial conditions used for a simulation, the distances between clusters was compared to these $r_{0}$ values. If a cluster was within $r_{0}$ of another, the pair's masses were combined and placed at their centre of mass, a new $r_{0}$ was calculated, and distances to other clusters checked again. The process repeated until no new clusters were added to the group, then started again using a different cluster as the starting point and repeated for all clusters. In all cases, except for SSC (due to the high population core), the process stopped after pairs were identified as bound.

In order to assess the state of the supercluster at the end of the simulations, kinetic and potential energies were compared. Since the velocities in the simulations were calculated relative to the centre of mass of the entire supercluster, the kinetic energy was calculated from velocities relative to the centre of mass of a particular pair or group of interest according to
\begin{equation}
\label{kinetic}
T = \sum_{i=1}^{N} \dfrac{1}{2} m_{i}(v_{i}-v_{cm})^{2},
\end{equation}
where $v_{cm}$ is the velocity of the centre of mass of the pair or group relative to the centre of mass of the supercluster. The potential energy was calculated as the normal Newtonian gravitational potential energy, and again focused on the particular pair or group of interest and not the entire supercluster,
\begin{equation}
\label{potential}
U = \sum_{i=1}^{N-1} \sum_{i<j}^{N} \dfrac{Gm_{i}m_{j}}{r_{ij}}.
\end{equation}

If the kinetic energy of the system was less than the potential energy, the group was considered bound. Given the dynamical time scales involved with superclusters, we would not expect any of the extended structures we find to have virialized by the end of a simulation. However, as long as the total kinetic energy of the system is lower than the potential energy of that system, it should remain bound.

In addition to the energies, close encounters were tracked during the course of the simulations. If the centres of a pair of clusters came within 3 Mpc of each other, the event was logged as a close encounter. Due to the simplicity of the model being used, with the clusters as point particles, tidal friction effects would be ignored. In larger $N$ simulations ($N \sim 3000$) we have performed using the same methodology as the smaller $N$ simulations described above, these effects will often quickly lead to mergers once clusters have come within 3 Mpc of each other. Thus, these events were also considered when making final conclusions about the structures in simulation.

\begin{figure}
\centering
\includegraphics[width=1\linewidth]{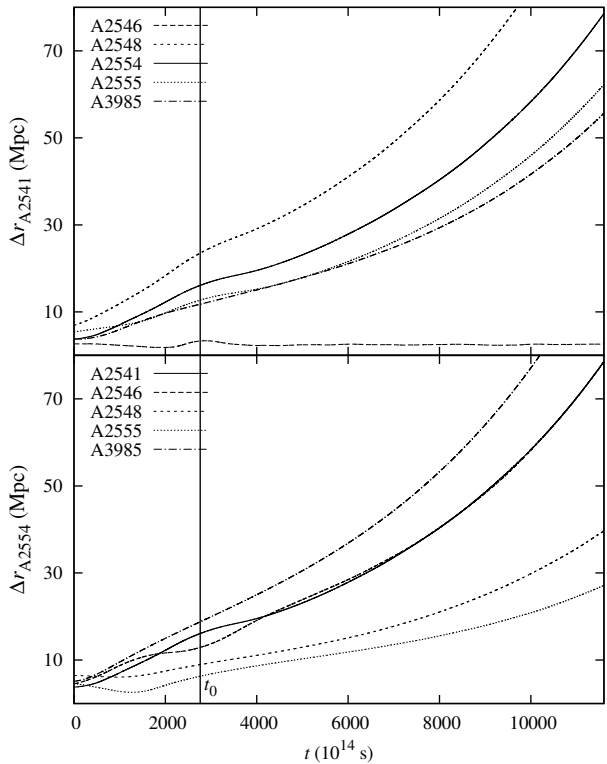}
\caption{Separations versus time averaged over all 250 simulation runs using the mass central values from Table \ref{ClusterProp} for ASC. The vertical line represents the current time, while $t = 0$ corresponds to $\sim 8.76 \, \mathrm{Gyr}$ in the past. The distances on the vertical axis assume an $h = 0.74$. The top panel shows positions relative to A2541, the bottom panel shows positions relative to A2554. Note that in the top panel A2546 does not expand away from A2541 like all other clusters. In the bottom panel A2555 is seen to expand away from A2554 but at a slower rate than the rest of the clusters.}
\label{AQSep}
\end{figure}

In order to give a rough estimate of the probability of the pairs or groups of clusters being bound, we simply take either the number of simulations in which a pair or group is found to be bound or has close encounters (which ever is larger) and divide it by the total number of simulations. We find that this simple calculation provides a good numerical representation of the likelihood of structures being bound and is supported by Figs. \ref{AQSep}--\ref{SCSep}, which are discussed in more detail in section 4.

Since we did not include any intercluster mass in our simulations, which may be part of the bound structure, and added an unperturbed Hubble expansion velocity to the initial conditions at the present, the results from this analysis should represent the minimum bound structure present in the superclusters. Future work will examine the effects of additional bound intercluster mass, as well as reductions in the Hubble expansion velocity added to the initial conditions (Pearson \& Batuski in preparation).
\section{Results}
\begin{table*}
\centering
\caption{Simulation Results}
\label{Results}
\begin{threeparttable}
\begin{tabular}{ccccccc}
\hline
\hline
\rule{0pt}{12pt}
Supercluster & Group & D06 & B03 & Results & Close Encounters & \% Likelihood \\ [1ex]
\hline
\rule{0pt}{12pt}
ASC & A2541/A2546 & $250 \pm 0.0$ & $246 \pm 1.8$ & $250 \pm 0.5$ & $249 \pm 0.2$ & $\sim100$ \\
${}$ & A2554/A2555 & $244 \pm 2.3$ & $162 \pm 14.3$ & $63 \pm 20.1$ & $54 \pm 17.2$ & 25.2 \\
${}$ & A2548/A2554 & $136 \pm 8.3$ & $72 \pm 8.6$ & $70 \pm 9.0$ & $35 \pm 6.7$ & 28.0 \\ [1ex]
\hline
\rule{0pt}{12pt}
MSC & A3695/A3696 & $224 \pm 59.2$ & $45 \pm 53.8$ & $191 \pm 63.7$ & $109 \pm 68.0$ & 76.4 \\
${}$ & A3693/A3695 & $47 \pm 30.1$ & $4 \pm 3.5$ & $5 \pm 17.2$ & $2 \pm 6.4$ & 2.0 \\
${}$ & A3693/A3696 & $35 \pm 23.5$ & $3 \pm 1.7$ & $2 \pm 9.8$ & $0 \pm 5.5$ & 0.8 \\
${}$ & A3693/A3705 & $69 \pm 23.4$ & $16 \pm 9.8$ & $2 \pm 6.3$ & $1 \pm 2.0$ & 0.8 \\ [1ex]
\hline
\rule{0pt}{12pt}
CBSC & A2061/A2067 & $68 \pm 19.8$ & $6 \pm 2.1$ & $22 \pm 5.9$ & $8 \pm 2.4$ & 8.8 \\
${}$ & A2065/A2089 & $40 \pm 27.4$ & $1 \pm 1.7$ & $7 \pm 7.5$ & $1 \pm 1.9$ & 2.8 \\ [1ex]
\hline
\rule{0pt}{12pt}
SSC & A3554/56/58/60/62 & $249 \pm 0.4$ & $146 \pm 14.8$ & $250 \pm 0.6$ & $205 \pm 8.3$\tnote{a} & $\sim100$ \\
${}$ & A3528/30/32 & $88 \pm 7.0$ & $15 \pm 2.1$ & $29 \pm 6.8$ & $5 \pm 1.8$\tnote{b} & 11.6 \\
${}$ & A3530/A3532 & $129 \pm 4.1$ & $54 \pm 3.7$ & $129 \pm 6.1$ & $91 \pm 5.8$ & 51.6 \\
${}$ & A1735/A3559 & $141 \pm 7.8$ & $72 \pm 6.5$ & $122 \pm 8.4$ & $116 \pm 5.2$ & 48.8 \\
${}$ & A3564/A3566 & $202 \pm 3.0$ & $161 \pm 3.2$ & $73 \pm 7.8$ & $90 \pm 2.7$ & 36.0 \\ [1ex]
\hline
\end{tabular}
\begin{tablenotes}
\item[a] This is the number of close encounters of A3554 with A3562, which is the smallest number of close encounters among the pairs in this group.
\item[b] This is the number of close encounters of A3528 with A3532, which is the smallest number of close encounters among the pairs in this group.
\end{tablenotes}
\end{threeparttable}
\end{table*}

Except for SSC, we did not find evidence for extended bound structures in these superclusters. By looking at the number of simulations in which pairs or groups are bound according to their energies, and the number of simulations in which there are close encounters, it was possible to conclude how likely it is that they are bound. In addition, it was possible to assess the predictive abilities of B03 and D06 in being applied to real structures in the universe, by comparing their predictions with the simulation results. 

Table \ref{Results} summarizes our findings using the central values of the masses listed in Table \ref{ClusterProp}. Listed are the pairs and groups of clusters that are likely bound, along with a few interesting cases discussed in further detail in the following paragraphs. Column 1 lists the supercluster membership of the pairs or groups. Column 2 lists the clusters in the pair or group. Columns 3 and 4 list the number of simulations in which the methods of D06 and B03 respectively, predict the pair or group as being bound based on the initial conditions. Column 5 lists the number of simulation runs in which the energies of the pair or group in the future frame would suggest that they are bound. Column 6 lists the number of simulations in which the pair or group has close encounters. Column 7 lists the per cent likelihood of the pair or group actually being bound. 

The listed dispersions come from the 20 Gaussian random mass realisations described in Section \ref{ICs}. By varying the mass, the value of $r_{0}$ from equations \eqref{r_B03} and \eqref{r_D06} will change, altering the predictions of the analytical models. In addition, different masses will give rise to different gravitational accelerations in the simulations, altering the outcomes. The results of each mass realisation are determined with the methods discussed in Section \ref{Analysis}. The dispersions are then the standard deviations of the results from each of the 20 mass realisations.  Given the small number of realisations for each supercluster, the dispersions listed in Table \ref{Results} may not be representative of the true values. However, they should give a sense of the effect the uncertainty in mass can have on the results. It should be noted that the structures we claim as having a strong likelihood ($\sim 100$ per cent) of being bound have very small dispersions. On the other hand, A3695 and A3696 in MSC do have large dispersions, though this pair does seem to represent a loosely bound structure (see the discussion below).

\subsection{The Aquarius Supercluster}
In ASC, there is evidence for a single pair of galaxy clusters being bound, A2541 and A2546. Looking at Table \ref{Results}, we can see that in all 250 simulations of the supercluster the pair is bound according to their energies in the future frame. They also have close encounters in 249 of the simulations. The analytical models do well at identifying this pair as bound. B03 predicts the pair to be bound, based on the initial conditions, in 246 of the simulations, while D06 predicts the pair to be bound in all 250 simulations. Given the fact that this pair is extremely likely to be gravitationally bound, there is not much room for the analytical models to be incorrect, so it is hard to draw a firm conclusion about their accuracy based on this pair.

Looking at a different pair in ASC, which seems not to be bound, it is possible to make an observation about the analytical models. A2554 and A2555 are predicted to be bound to each other in a significant fraction of the simulations by both B03 and D06, 162 and 244 respectively, yet they are bound in far fewer simulations, and have even fewer closer encounters. The predictions of D06 are well separated from the results, even at the 3$\sigma$ level, and there is only slight overlap between the predictions of B03 and the results at the 3$\sigma$ level. The reason for the discrepancy in the models and the results can be found by looking at Fig. \ref{SCSky}. A2555 is located between A2554 and A2541/A2546, and in fact, in most of our simulations is just outside the combined $r_{0}$ of the pair, as calculated by D06. This proximity is what prevents A2555 from being bound to A2554 in most simulations, and shows an inherent flaw of the analytical models which was also noted by D06. Since both are based on the spherical collapse model, they assume that the fate of a test particle is determined solely by the mass enclosed by a particular shell. The interaction seen here shows that in dense environments, other nearby structures can play an important role.

Fig. \ref{AQSep} looks at the separation of the clusters in ASC averaged over all 250 simulations based on the mass central values. The top panel shows the separations relative to A2541. On average, A2546 remains very close ($\sim 3 \, \mathrm{Mpc}$) to A2541, strongly supporting our conclusion of this being a bound structure. The bottom panel shows the separations relative to A2554. Here it can be seen that the general trend is for all clusters to move away over time. Just after the current time, denoted by the vertical line, it appears that A2555 is still being slowed by the gravitational attraction to A2554,  but around time step 6000 dark energy clearly begins to dominate. The two different views help to get the full picture of the dynamics in the supercluster. In the top panel it seems as though A2555 and A3985 may be bound, but in the bottom panel it becomes clear this is not the case, given their drastically different separations from A2554. For the case of A2541 and A2546 we can see that in both panels they remain close together, clearly indicating that they are bound.

No other pairs or groups in ASC have a significant chance of being bound according to the simulations or the analytical models.
\subsection{The Microscopium Supercluster}
\begin{figure}
\centering
\includegraphics[width=1\linewidth]{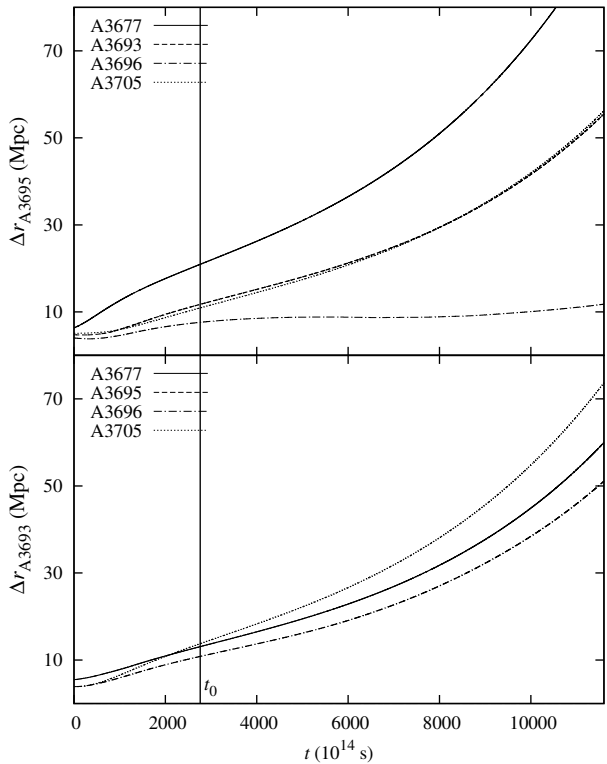}
\caption{Same as Fig. \ref{AQSep} except for MSC. The top panel shows positions relative to A3695, the bottom panel shows positions relative to A3693. Note that in the top panel A3696 remains, on average, within $\sim 10 \, \mathrm{Mpc}$ of A3695 far into the future. In the bottom panel it can be seen that all clusters seem to expand away from A3693.}
\label{MSSep}
\end{figure}
Similar to what was seen in ASC, MSC has a pair of clusters with a significant chance of being bound, A3695 and A3696 (see Table \ref{Results}). The pair seems to be close to the limit of being bound, given the comparatively small number of close encounters for the number of times they are bound according to their energies. It is also interesting to note that B03 would only predict the pair to be bound in 45 of the simulations, substantially below the number in which they end up being bound. D06, on the other hand, predicts them to be bound in 224 simulations, a good deal more than the 191 in which they are bound. Given the large dispersions for the results of this pair, it should be noted that the predictions of D06 are well within one standard deviation, while the predictions of B03 are further removed from the results, but would be consistent with the number of close encounters. 

Fig. \ref{MSSep} is similar to Fig. \ref{AQSep}, but for MSC. The trends in the cluster separations support our conclusions about the structure that is likely to be bound in this supercluster. In the top panel of Fig. \ref{MSSep} we can see that A3695 and A3696, on average, remain close to each other ($\sim 10 \, \mathrm{Mpc}$) far into the future, hinting that they are indeed likely to be bound. The slight upturn in the line for A3696 around time step 10000 shows the influence of the simulations in which the pair is not bound on the average. In the top panel it also seems that perhaps A3693 and A3705 may be bound to each other, given that they expand away from A3695 in a very similar manner. The bottom panel, looking at positions relative to A3693, shows a different story, with A3705 clearly expanding away.

Aside from A3695 and A3696, no other pairs or groups in MSC have a significant chance of being bound according to the simulations or the analytical models.
\subsection{The Corona Borealis Supercluster}
\begin{figure}
\centering
\includegraphics[width=1\linewidth]{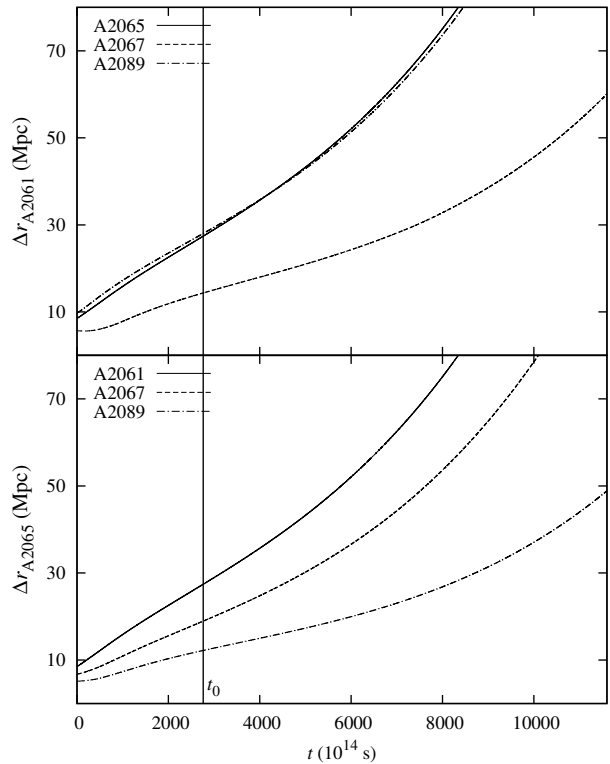}
\caption{Same as Fig. \ref{AQSep} except for CBSC. The top panel shows positions relative to A2061. The bottom panel shows positions relative to A2065. In both panels it is clear that the average trend is for the supercluster to expand.}
\label{CBSep}
\end{figure}
CBSC is interesting given the lack of evidence for any bound structure. The two entries in Table \ref{Results} are the pairs most likely to be bound in CBSC, and they do not have a significant chance of being bound. However, the pattern of the predictions of the analytical models is still seen. B03 predicts the clusters to be bound in fewer simulations than those in which they are found to be bound, while D06 predicts them to be bound in more simulations than those in which they are found to be bound, both being more than one standard deviation away from the results. Looking at Fig. \ref{CBSep}, we can see that the overall trend is for the supercluster to expand, supporting our conclusions about the structure.

The reason that the lack of bound structure is interesting is that other authors \citep{Small98,Kopylova98} have found observational evidence that this is a collapsing, gravitationally bound supercluster. The implications of this discrepancy are discussed further in section 5.1 of this paper.
\subsection{The Shapley Supercluster}
\begin{figure}
\centering
\includegraphics[width=1\linewidth]{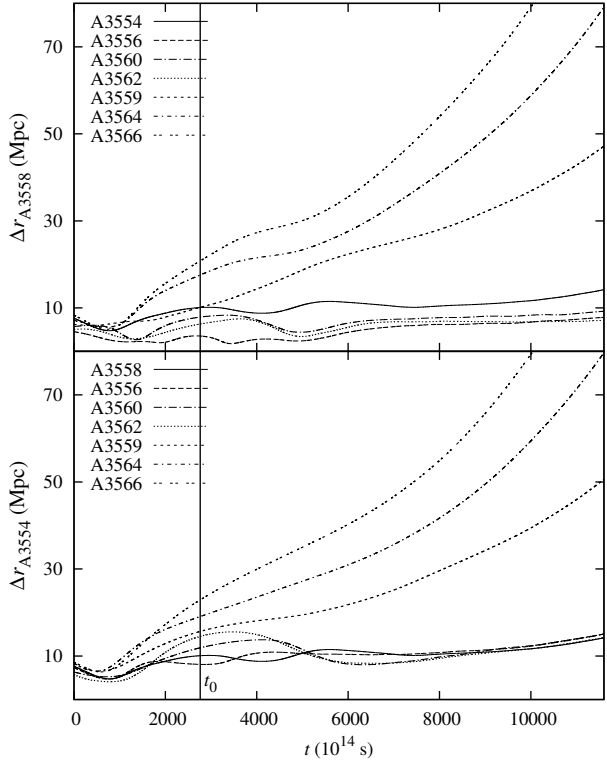}
\caption{Same as Fig. \ref{AQSep} except for SSC. The top panel shows positions relative to A3558, often taken to represent the core of the supercluster. The bottom panel shows positions relative to A3554, the most loosely bound member of the core. In both panels it can be seen that the five clusters making up the core, A3554, A3556, A3558, A3560, and A3562, all remain within $\sim 14 \, \mathrm{Mpc}$ of each other far into the future, indicating a strong likelihood of being bound. The other clusters plotted were chosen due to their proximity to the core (see Fig. \ref{SCSky}). They show signs of expansion, limiting the likelihood they are part of the bound structure.}
\label{SCSep}
\end{figure}
SSC is the largest collection of galaxy clusters in the local universe, representing an extreme overdensity. Many authors have studied this supercluster and have found some interesting results, including the possibility of the SSC significantly contributing to our Local Group's peculiar motion \citep{Proust06}. It is no surprise that this supercluster stands out as the only one to have a large group of clusters that seem to be bound. A3554, A3556, A3558, A3560, and A3562 have large numbers of close encounters during the simulations. A3554 and A3562, the most widely separated pair, have close encounters in 205 simulations. Looking at the other clusters, A3556, A3558, and A3560 have close encounters with A3562 in 250, 249, and 248 of the simulations respectively. Given that these close encounters tend to lead to mergers between the clusters, this group is likely bound. This is supported by their energies, which suggest that all five are bound in all 250 of the simulations. Upon visual examination of this group in the future frame of a good number of simulations, the structure that forms is rather planar. This is to be expected given the small number of particles, lack of mergers, and conservation of angular momentum.

Fig. \ref{SCSep} shows the relative positions of some of the clusters versus time, averaged over all 250 simulations of the supercluster. All five of the clusters that make up the core remain, on average, within $\sim 14 \, \mathrm{Mpc}$ of each other far into the future. The upturn of the line for A3554 around time step 8000 shows that it is the least likely of the five to remain bound. The fact that it is still, on average, very close to all the other clusters in the core at the end of the simulations, hints it still has a strong likelihood of being bound.

In addition to the group of five, there are a couple of other interesting pairs. A1736 and A3559 have some chance of being bound to each other, and the same pattern in regards to the analytical model predictions is observed. It should be noted that the predictions of D06 are consistent with the results at the 2$\sigma$ level, while those of B03 are not consistent with the results even at the 3$\sigma$ level. A3564 and A3566 are similar to what was seen with A2554 and A2555 in ASC. The analytical predictions hint that A3564 and A3566 should be bound, while the energies and number of close encounters hint otherwise. This is again due to the tidal effect of other nearby structure, namely the group of five clusters.
\section{Discussion}
Since our analysis should have found the minimum bound structure present in these superclusters, we can conclude that there is a good chance of seeing extended bound structure in both ASC and SSC, while there is also some chance of finding extended bound structure in MSC. The fact that the lower limit of bound structure in ASC and MSC seems to be a single pair of clusters, hints that there may in fact be between three and five clusters in an extended bound structure. SSC, having a lower limit of five bound clusters, raises the possibility that a substantial number of the clusters listed in Table \ref{ClusterProp} may actually be part of an extended bound structure.

Neither analytical model is completely consistent with the simulation results, especially given what is observed with SSC, which are our most reliable results. The model of B03 will pick out all structure that is definitely bound, but given the tendency to under-predict seen in the above, it may miss some structure that will in fact be bound. The model of D06 will identify all structure that is bound but, given the tendency to over-predict, it may falsely identify some structures as bound. The model of D06 may have performed better had we not assumed an unperturbed Hubble expansion velocity at present. Neither model does well if there are other structures nearby.
\subsection{Incomplete, Incorrect or Both?}
If the superclusters are actually gravitationally bound, as some authors claim for the case of CBSC, that means one of three things, 1) our model of the superclusters is incomplete, 2) our model of the superclusters is incorrect, or 3) both. To examine this in some detail, consider the work of \cite{Small98}. In the end they claim the mass of the supercluster is $\sim 4 \times 10^{16} \, h^{-1} \, \mathrm{M}_{\odot}$, while the sum of the masses of the clusters they include in the supercluster (A2061, A2065, A2067, A2079, A2089, A2092 and Cl1529+29) is only $5.3 \times 10^{15} \, h^{-1} \, \mathrm{M}_{\odot}$. This means there would have to be a substantial mass component outside of the clusters themselves, about seven times the mass of the clusters, in order to make up the difference. This amount of mass if far more than the amount found to be outside the clusters in SSC, which may have twice as much mass in the intercluster galaxies as in the cluster galaxies \citep{Proust06}. At first glance this seems like an unreasonably large amount of mass, but taking the volume they claim for the supercluster, $2.8 \times 10^{4} \, f \, h^{-3} \, \mathrm{Mpc}^{3}$, where $f$ is the dimensionless elongation parameter relating the elongation along the line of sight to that on the sky, it's possible to determine the density needed to reach this mass. Using an $f$ value of 5, which indicates that the diameter along the line of sight is actually about the same as the diameter on the sky (i.e. roughly spherical), the average spatial density to get to the mass claimed would be $\sim 3 \times 10^{11} \, h^{2} \, \mathrm{M}_{\odot} \, \mathrm{Mpc}^{-3}$. As reference, the clusters in CBSC have densities of $\sim 10^{15} \, h^{2} \, \mathrm{M}_{\odot} \, \mathrm{Mpc}^{-3}$, and the critical density is $\sim 2.8 \times 10^{11} \, h^{2} \, \mathrm{M}_{\odot} \, \mathrm{Mpc}^{-3}$. The fact that the density needed to get to the mass claimed is essentially the same as the critical density is rather coincidental and may imply that their methods are simply generating the mass required to bind the supercluster. Therefore, the results of \cite{Small98} on their own, are not all that compelling.

On the other hand, there is some observational evidence that the core of CBSC may be collapsing. \cite{Kopylova98} used the Kormendy relation \citep{Kormendy77} to determine redshift independent distances to the clusters in CBSC. With this information, they could then determine whether the clusters had velocities towards or away from the supercluster centroid. They claim that the core consisting of A2061, A2065, A2067, A2089 and A2092 shows evidence of being in a state of collapse. More recently Batiste \& Batuski (2013, in preparation) have found similar results by employing the fundamental plane relation \citep{Djorgovski87} which can determine distances more precisely than the Kormendy relation. This observational evidence makes the case for a large matter component outside of the clusters more compelling.

The lack of knowledge about the true peculiar velocities at the present time may also influence our results. For all of the clusters studied here, the data available are the positions on the sky and the spectroscopic redshifts, which are affected by the peculiar velocities. If there was enough data to determine redshift independent distances for the clusters, similar to what has been doarXivne with CBSC, it would be possible to constrain the peculiar velocities and better sort out the three dimensional positions of the clusters. Looking at Fig. \ref{SCSky}, it is clear why this may be important. In all four panels there are pairs of clusters that are very close to each other on the sky. If they are indeed close together in real space, their peculiar velocities could influence the measured spectroscopic redshift, which could substantially alter the model of the supercluster.

The other reason for disagreement between the simulations and the observations may have to do with incorrect assumptions made in determining initial conditions. The most likely culprit would be the addition of the Hubble flow component to the velocities. If the clusters have slowed and begun to break away from the Hubble flow, this would increase the likelihood of the clusters being bound. This matter is currently under investigation in a new round of simulations (Pearson \& Batuski in preparation). If the results simply suggest that the clusters must have slowed relative to the Hubble flow by some amount, then there may be no need to assume there is a substantial intercluster matter component.

Lastly, it is possible that the current observational data is not providing the full picture of these superclusters and the assumptions made are incorrect. For two of the superclusters, ASC and MSC, several of the mass estimates are based on fewer than ten measured redshifts in the cluster. This calls into question the accuracy of the calculated dispersions, and it is possible that with more redshifts the mass estimate would be revised upward, though likely not enough to bind the superclusters. However, combining this with a reduction of Hubble flow velocity may change the results of our work.
\subsection{The Analytical Models}
In the end, neither analytical model seemed to be able to consistently identify real bound structures in simulation. The model of B03 assumed that test particles are currently streaming away from structures with an unperturbed Hubble flow velocity. This is similar to the assumption made in the initial conditions of the simulations presented here, and therefore should be an accurate predictor of bound structure in those simulations. In fact, since we also included a tangential component to the velocity, if anything the model of B03 should predict more structure as being bound than we find. However, the results suggest that the model of B03 is still likely to miss some bound structure.

The model of D06 makes no assumptions about the current velocities of test particles. Instead, they determine that the energy of a critically bound shell at present must be $\widetilde{E} = 3/2$ in dimensionless units. With this, they then integrate the energy equation of the spherical collapse model to find the radius of the shell associated with that energy at the present time. This becomes the basis for equation \eqref{r_D06}. They also determine that a critical shell should have slowed to about 29 per cent of the Hubble flow velocity at the present time. Thus, if these structures are indeed bound, they will have at least slowed, if not already broken from the Hubble flow. As discussed above, this appears to be the case for CBSC. Future simulations will be aimed at determining the effects of reducing the Hubble flow component, or allowing for the possibility of some inward component of initial velocity. Despite avoiding assumptions about current velocities, the model of D06 will falsely identify some structures as bound. The authors themselves note this flaw, attributing it to ignoring the effects of tangential motions of test particles and external attractors. \cite{Dunner07}, when extending the results to redshift space, found that the criterion performed better than in real space. Unfortunately, the extension to redshift space would require simulations performed with large numbers of particles, making it impractical to apply the results of \cite{Dunner07} to our simulations.

Since both analytical models are based on the spherical collapse model, which only considers the mass within a shell, neither is reliable when other structures are nearby. In more isolated regions, the models of B03 and D06 can be used to give lower and upper limits to the extent of bound structures, respectively.

${}$

In the end, further study of these candidates is needed. In both ASC and MSC, more large area spectroscopy studies should be performed to increase the number of measured redshifts so that the masses of the clusters can be better constrained. By examining the line widths of the spectra and the photometric properties, it would also be possible to apply the Kormendy relation and fundamental plane to obtain secondary distance indicators. If photometric surveys revealed a large enough sample of edge-on spiral galaxies, it may also be possible to make use of the Tully-Fisher relation for a secondary distance indicator. These could then be used to place constraints on the peculiar velocities of the clusters. If a large enough number of field galaxies in the region are also surveyed an analysis similar to that of \cite{Small98} can be performed, possibly placing an upper limit on the total mass in the region, including the mass outside of the clusters. Weak lensing studies would be an additional method to constrain the amount of mass in the region of these superclusters, possibly answering the question of whether there is significant dark matter in the intercluster region.
\section*{Acknowledgments}
The authors wish to thank Merida Batiste for private discussions of her results in regards to CBSC. This research was partially supported by grants from NASA through the Maine Space Grant Consortium (MSGC). This research also made use of SAO/NASA Astrophysics Data System (ADS) and the National Virtual Observatory (NVO).
\bibliographystyle{mn2e}
\bibliography{LBSAU-Paper}

\begin{thebibliography}{}

\bibitem[\protect\citeauthoryear{Abell, Corwin Harold~G. \& Olowin}{Abell
  et~al.}{1989}]{ACO}
Abell G.~O.,  Corwin Harold~G. J.,    Olowin R.~P.,  1989, ApJS, 70, 1

\bibitem[\protect\citeauthoryear{Ade, Aghanim, Armitage-Caplan \& et al.}{Ade
  et~al.}{2013}]{Ade13}
Ade P. A.~R.,  Aghanim N.,  Armitage-Caplan C.,    et al. 2013, preprint
  (arXiv:1303.5076)

\bibitem[\protect\citeauthoryear{Amanullah, Lidman, Rubin \& et al.}{Amanullah
  et~al.}{2010}]{Amanullah10}
Amanullah R.,  Lidman C.,  Rubin D.,    et al. 2010, ApJ, 716, 712

\bibitem[\protect\citeauthoryear{Baryshev, Chernin \& Teerikorpi}{Baryshev
  et~al.}{2001}]{Baryshev01}
Baryshev Y.~V.,  Chernin A.~D.,    Teerikorpi P.,  2001, A{\&}A, 378, 729

\bibitem[\protect\citeauthoryear{Batuski, Miller, Slinglend \& et al.}{Batuski
  et~al.}{1999}]{Batuski99}
Batuski D.~J.,  Miller C.~J.,  Slinglend K.~A.,    et al. 1999, ApJ, 520, 491

\bibitem[\protect\citeauthoryear{Beers, Flynn \& Gebhardt}{Beers
  et~al.}{1990}]{Beers90}
Beers T.~C.,  Flynn K.,    Gebhardt K.,  1990, AJ, 100, 32

\bibitem[\protect\citeauthoryear{Busha, Adams, Wechsler \& Evrard}{Busha
  et~al.}{2003}]{Busha03}
Busha M.~T.,  Adams F.~C.,  Wechsler R.~H.,    Evrard A.~E.,  2003, ApJ, 596,
  713

\bibitem[\protect\citeauthoryear{Caretta, Maia, Kawasaki \& Willmer}{Caretta
  et~al.}{2002}]{Caretta02}
Caretta C.~A.,  Maia M. A.~G.,  Kawasaki W.,    Willmer C. N.~A.,  2002, AJ,
  123, 1200

\bibitem[\protect\citeauthoryear{Caretta, Maia \& Willmer}{Caretta
  et~al.}{2004}]{Caretta04}
Caretta C.~A.,  Maia M. A.~G.,    Willmer C. N.~A.,  2004, AJ, 128, 2642

\bibitem[\protect\citeauthoryear{Carlberg, Yee, Ellignson \& et al.}{Carlberg
  et~al.}{1996}]{Carlberg96}
Carlberg R.~G.,  Yee H. K.~C.,  Ellignson E.,    et al. 1996, ApJ, 462, 32

\bibitem[\protect\citeauthoryear{Colless, Peterson, Jackson \& et al.}{Colless
  et~al.}{2003}]{2dF}
Colless M.,  Peterson B.~A.,  Jackson C.,    et al. 2003, preprint
  (arXiv:astro-ph/0306581)

\bibitem[\protect\citeauthoryear{Da~Costa, Willmer, Pellegrini \& et
  al.}{Da~Costa et~al.}{1998}]{DaCosta98}
Da~Costa L.~N.,  Willmer C. N.~A.,  Pellegrini P.~S.,    et al. 1998, AJ, 116,
  1

\bibitem[\protect\citeauthoryear{Dietrich, Werner, Clowe \& et al.}{Dietrich
  et~al.}{2012}]{Dietrich12}
Dietrich J.~P.,  Werner N.,  Clowe D.,    et al. 2012, Nat, 487, 202

\bibitem[\protect\citeauthoryear{Djorgovski \& Davis}{Djorgovski \&
  Davis}{1987}]{Djorgovski87}
Djorgovski S.,  Davis M.,  1987, ApJ, 313, 59

\bibitem[\protect\citeauthoryear{Dressler}{Dressler}{1991}]{Dressler91}
Dressler A.,  1991, ApJS, 75, 241

\bibitem[\protect\citeauthoryear{D{\"{u}}nner, Araya, Meza \&
  Reisenegger}{D{\"{u}}nner et~al.}{2006}]{Dunner06}
D{\"{u}}nner R.,  Araya P.~A.,  Meza A.,    Reisenegger A.,  2006, MNRAS, 366,
  803

\bibitem[\protect\citeauthoryear{D{\"{u}}nner, Reisenegger, Meza \&
  Araya}{D{\"{u}}nner et~al.}{2007}]{Dunner07}
D{\"{u}}nner R.,  Reisenegger A.,  Meza A.,    Araya P.~A.,  2007, MNRAS, 376,
  1577

\bibitem[\protect\citeauthoryear{Fairall \& Jones}{Fairall \&
  Jones}{1991}]{Fairall91}
Fairall A.~P.,  Jones A.,  1991, Southern Redshift Catalogue and Plots.
Publications of the Dept. of Astron., Univ of Cape Town, N.11, South Africa

\bibitem[\protect\citeauthoryear{Gray, Taylor, Meisenheimer \& et al.}{Gray
  et~al.}{2002}]{Gray02}
Gray M.~E.,  Taylor A.~N.,  Meisenheimer K.,    et al. 2002, ApJ, 568, 141

\bibitem[\protect\citeauthoryear{Guzzo, Schuecker, B{\"{o}}hringer \& et
  al.}{Guzzo et~al.}{2009}]{Guzzo09}
Guzzo L.,  Schuecker P.,  B{\"{o}}hringer H.,    et al. 2009, A{\&}A, 499, 357

\bibitem[\protect\citeauthoryear{Hanami, Tsuru, Shimasaku \& et al.}{Hanami
  et~al.}{1999}]{Hanami99}
Hanami H.,  Tsuru T.,  Shimasaku K.,    et al. 1999, ApJ, 521, 90

\bibitem[\protect\citeauthoryear{Hernquist \& Katz}{Hernquist \&
  Katz}{1989}]{Hernquist89}
Hernquist L.,  Katz N.,  1989, ApJS, 70, 419

\bibitem[\protect\citeauthoryear{Heymans, Gray, Peng \& et al.}{Heymans
  et~al.}{2008}]{Heymans08}
Heymans C.,  Gray M.~E.,  Peng C.~Y.,    et al. 2008, MNRAS, 385, 1431

\bibitem[\protect\citeauthoryear{Huchra, Davis, Latham \& Tonry}{Huchra
  et~al.}{1983}]{CfA}
Huchra J.,  Davis M.,  Latham D.,    Tonry J.,  1983, ApJS, 52, 89

\bibitem[\protect\citeauthoryear{Huchra, Macri, Masters \& et al}{Huchra
  et~al.}{2012}]{Huchra12}
Huchra J.~P.,  Macri L.~M.,  Masters L.~M.,    et al 2012, ApJS, 199, 26

\bibitem[\protect\citeauthoryear{Jones, Read, Saunders \& et al.}{Jones
  et~al.}{2009}]{6dF}
Jones D.~H.,  Read M.~A.,  Saunders W.,    et al. 2009, MNRAS, 399, 683

\bibitem[\protect\citeauthoryear{Katgert, Mazure, Perea \& et al.}{Katgert
  et~al.}{1996}]{Katgert96}
Katgert P.,  Mazure A.,  Perea J.,    et al. 1996, A{\&}A, 310, 8

\bibitem[\protect\citeauthoryear{Kopylova \& Kopylov}{Kopylova \&
  Kopylov}{1998}]{Kopylova98}
Kopylova F.~G.,  Kopylov A.~I.,  1998, AstL, 24, 491

\bibitem[\protect\citeauthoryear{Kormendy}{Kormendy}{1977}]{Kormendy77}
Kormendy J.,  1977, ApJ, 218, 333

\bibitem[\protect\citeauthoryear{Lavaux \& Hudson}{Lavaux \&
  Hudson}{2011}]{Lavaux11}
Lavaux G.,  Hudson M.~J.,  2011, MNRAS, 416, 2840

\bibitem[\protect\citeauthoryear{Miller, Slinglend, Batuski \& Hill}{Miller
  et~al.}{1999}]{Miller99}
Miller C.,  Slinglend K.,  Batuski D.,    Hill J.,  1999, ApJ, 523, 492

\bibitem[\protect\citeauthoryear{Perlmutter, Aldering, Goldhaber \& et
  al.}{Perlmutter et~al.}{1999}]{Perlmutter99}
Perlmutter S.,  Aldering G.,  Goldhaber G.,    et al. 1999, ApJ, 517, 565

\bibitem[\protect\citeauthoryear{Postman, Geller \& Huchra}{Postman
  et~al.}{1988}]{Postman88}
Postman M.,  Geller M.~J.,    Huchra J.~P.,  1988, AJ, 95, 267

\bibitem[\protect\citeauthoryear{Proust, Quintana, Carrasco \& et al.}{Proust
  et~al.}{2006}]{Proust06}
Proust D.,  Quintana H.,  Carrasco E.,    et al. 2006, A\&A, 447, 133

\bibitem[\protect\citeauthoryear{Reisenegger, Quintana, Carrasco \&
  Maze}{Reisenegger et~al.}{2000}]{Reisenegger00}
Reisenegger A.,  Quintana H.,  Carrasco E.~R.,    Maze J.,  2000, AJ, 120, 523

\bibitem[\protect\citeauthoryear{Ryden}{Ryden}{2003}]{Ryden03}
Ryden B.,  2003, Introduction to Cosmology.
Addison Wesley, San Francisco, CA

\bibitem[\protect\citeauthoryear{Sandage}{Sandage}{1975}]{Sandage75}
Sandage A.,  1975, ApJ, 202, 563

\bibitem[\protect\citeauthoryear{Small, Ma, Sargent \& Hamilton}{Small
  et~al.}{1998}]{Small98}
Small T.~A.,  Ma C.-P.,  Sargent W. L.~W.,    Hamilton D.,  1998, ApJ, 492, 45

\bibitem[\protect\citeauthoryear{Small, Sargent \& Hamilton}{Small
  et~al.}{1997}]{Small97}
Small T.~A.,  Sargent W. L.~W.,    Hamilton D.,  1997, ApJS, 111, 1

\bibitem[\protect\citeauthoryear{Spergel, Verde, Peiris \& et al.}{Spergel
  et~al.}{2003}]{Spergel03}
Spergel D.~N.,  Verde L.,  Peiris H.~V.,    et al. 2003, ApJS, 148, 175

\bibitem[\protect\citeauthoryear{Springel}{Springel}{2005}]{Springel05}
Springel V.,  2005, MNRAS, 364, 1105

\bibitem[\protect\citeauthoryear{Struble \& Rood}{Struble \&
  Rood}{1999}]{Struble99}
Struble M.,  Rood H.,  1999, ApJS, 125, 35

\bibitem[\protect\citeauthoryear{Wang \& Rowan-Robinson}{Wang \&
  Rowan-Robinson}{2009}]{Wang09}
Wang L.,  Rowan-Robinson M.,  2009, MNRAS, 398, 109

\bibitem[\protect\citeauthoryear{Zucca, Zamorani, Scaramella \&
  Vettolani}{Zucca et~al.}{1993}]{Zucca93}
Zucca E.,  Zamorani G.,  Scaramella R.,    Vettolani G.,  1993, ApJ, 407, 470

\end{thebibliography}

\label{lastpage}

\end{document}